\documentstyle{article}

\begin{document}

\begin{titlepage}

\begin{flushleft}
TU-530
\end{flushleft}

\begin{center}
{\LARGE \bf
 Dynamical Instability of a two-dimensional Quantum Black Hole}
\end{center}

\begin{center}
\Large{
 Y. Itoh${\ }^{a}$, M. Hotta${\ }^{b}$, M. Morikawa${\ }^{c}$ and T. 
Futamase${\ }^{a}$
 }\\

{\it 
${\ }^{a}$ Astronomical Institute, Faculty of Science, Tohoku University, 
Sendai 980-77, Japan \\
${\ }^{b}$ Department of Physics, Faculty of Science, Tohoku University,\\
Sendai 980-77, Japan\\
${\ }^{c}$ Department of Physics, Faculty of Science, Ochanomizu
Women's  University,\\
Tokyo, Japan\\
} 
\end{center}

\begin{abstract}
          
We investigate dynamical instability of a two-dimensional quantum 
black hole model considered by Lowe in his study of Hawking evaporation.  
The model is supposed to express a black hole in equilibrium with a bath 
of Hawking radiation. It turns out that the model has at least one 
instability modes for a wide range of parameters, and thus it 
is unstable. 

\end{abstract}

\end{titlepage}

\section{Introduction}

The final stage of black hole evaporation and associated problem have 
attracted much theoretical attention since the evaporation process is 
discovered by Hawking\cite{H}. 
Although various suggestions have been proposed in the past, 
serious theoretical treatments have started in 1992 by Callan, Giddings, 
Harvey and Strominger(CGHS) \cite{CGHS} using a toy model in two-dimensional dilaton gravity coupled with conformal matter. 
They have shown that their model(hence force called CGHS model) is able to describe the formation of a black hole and Hawking radiation. Moreover 
the backreaction of the metric is analyzed to leading order in a 1/N 
expansion, where N is the number of matter fields. 
 Static solutions are also found in the CGHS model which is supposed 
to describe a black hole in equilibrium with a bath of Hawking radiation\cite{BGHS}. 

Stimulated by their work, various models of two-dimensional dilaton gravity 
are investigated\cite{DG,H92,ST}. 
Among them Lowe studied an interesting model which 
is obtained by the dimensional reduction of the four-dimensional Einstein action\cite{L}. 
In his treatment the two-dimensional matter is coupled to the model 
to allow the backreaction in the same way as the CGHS model. 
Lowe has also obtained static models which correspond to infinite 
mass black holes supported by an incoming flux of radiation just like 
the static solution of the CGHS model.  
These static solutions are of some interest because the evaporation of 
a large mass black hole may be well approximated by a succession of these  
solutions.   

In spite of their importance in understanding the evaporation process, 
the stability analysis of these static solutions in the CGHS model 
as well as Lowe's model have not been made. We mean stability 
in the following sense. The static solutions mentioned above are 
actually the background configuration in a pure state defined 
by the effective action. If one would like to think of a quantum black hole 
in equilibrium with a thermal bath, one has to take the thermal state 
and thermal fluctuations will inevitably present. 
Then the question arises naturally if the static black holes  
are stable against such fluctuations. 
The aim of the present paper is to address this question, and to perform  
a stability analysis for the quantum black hole. 
Two-dimension will be an unique place to be able to do it 
in an explicit way. Here we shall take Lowe's model as our example. 
Our method is completely general and applicable to any models.  
It is also important to realize that the instability discussed 
here has nothing to do with  thermal instability due to the  negative heat capacity in the case of the four-dimensional 
evaporating black hole in a thermal bath\cite{D}. 
In fact the Hawking temperature of the two-dimensional black hole 
is independent of the mass, and thus there will be no possibility to have 
thermal instability in our case. 

This paper is organized as follows. First we review the static solutions in Lowe's model in section 2. Then we shall derive linear perturbation around 
the static solutions in section 3. 
There we shall show that the perturbed variable obeys the wave equation 
with potential. We also show that the potential is negative near 
the horizon.
 In section 4 we also perform numerical 
analysis to confirm that the potential is negative definite and 
there will be at least one unstable mode. Some discussions are 
devoted in section 5.

\section{Static solutions in Lowe's model}

We shall first review Lowe's model and its static solutions. His model 
is obtained by the dimensional reduction of the four-dimensional Einstein gravity. Here we shall consider 
the following action. 
\begin{eqnarray}
    S &=&        \frac{1}{2\pi} \int d^2 x \sqrt{-g} 
                 \left( e^{-2\phi} R + 2e^{-2\phi}(\nabla \phi)^2 
               + 4\lambda^2
               - \frac{1}{2}(\nabla f)^2 \right)   \nonumber \\
    \mbox{}  &-& \frac{\kappa}{8\pi}\int d^2 x \sqrt{-g} R\Box^{-1}R .   
\label{action}  
\end{eqnarray}
where the coupling constant $\kappa$ is determined by the number of matter fields $N$ as
\begin{eqnarray}
    \kappa  &=& \frac{N}{12} .
\end{eqnarray}
The above action reduces to  Lowe's action when 
$\lambda = \frac{1}{\sqrt{2}}$. We shall not specify  any particular value 
for $\lambda$ in our analysis below.

We shall take the following form of the metric
\begin{eqnarray}
    ds^2    &=&    -e^{2\rho}dx^{+}dx^{-} .
\label{metric}
\end{eqnarray}
Then the equation of motion may be written as
\begin{eqnarray}
    \partial_{+}\partial_{-}\phi     &=& 
         \partial_{+}\phi\partial_{-}\phi  
                                                    \nonumber  \\
    \mbox{}                          &+&  
         \frac{1}{1 - \frac{\kappa}{2}e^{2\phi}}
         (\partial_{+}\phi\partial_{-}\phi 
     +  {\textstyle  \frac{\lambda^{2}}{2}} e^{2(\rho+\phi)}) \label{em1} 
                                                                 \\
    \partial_{+}\partial_{-}\rho     &=& 
         \partial_{+}\partial_{-}\phi 
      -  \partial_{+}\phi\partial_{-}\phi      \label{em2}
                                                                  \\
    \partial_{+}\partial_{-}f        &=&  0 .\label{em3}
\end{eqnarray}     
The constraint equations are 
\begin{eqnarray}
0  =  T_{\pm\pm} 
  &=& e^{-2\phi}[  4\partial_{\pm}\rho\partial_{\pm}\phi
       - 2\partial^{2}_{\pm}\phi
       + 2(\partial_{\pm}\phi)^{2}]
       +  {\textstyle \frac{1}{2}}(\partial_{\pm}f)^{2}  \nonumber \\
 \mbox{}   &-& 
         \kappa[  (\partial_{\pm}\rho)^{2} 
       - \partial^2_{\pm}\rho
       + t_{\pm}(x^{\pm}) ] . \label{cnst} 
\end{eqnarray}
The equations (\ref{em1}) and (\ref{em2}) shows that the dilaton field takes the following value at singularity.
\begin{eqnarray}
     \phi_{cr} &=& - \frac{1}{2}\ln (\frac{\kappa}{2}) .
\end{eqnarray}   
In the following we shall use the new coordinates defined by 
\begin{eqnarray*}
     \sigma     &=&    {\textstyle \frac{1}{2} } (x^{+} - x^{-}) .  \\
     t          &=& {  \textstyle \frac{1}{2} } (x^{+} + x^{-}) .  \\
\end{eqnarray*}
In this coordinate horizon is at $\sigma = -\infty$.
 
In this paper we are interested in the static solution of the above set of equations which corresponds to a black hole in equilibrium with Hawking radiation. The above equations of motion(\ref{em1}-\ref{em3}) take the following forms in the static situation in the new coordinates;   
\begin{eqnarray}
\phi'' &=& 
       (\phi')^{2} + \frac{1}{1 - \frac{\kappa}{2}e^{2\phi}}
       ((\phi')^{2} - 2\lambda^{2} e^{2(\rho+\phi)}) \label{sem1}  \\
\rho''&=& 
       \phi'' - (\phi')^{2} . \label{sem2} \\
 f &=& 0 .  \label{sem3}
\end{eqnarray}
Also the constraint equation(\ref{cnst}) becomes
\begin{eqnarray}
    \rho'\phi' - {\textstyle \frac{1}{2}}\phi'' 
  + {\textstyle \frac{1}{2}}(\phi')^{2}      
 &=&
    {\textstyle \frac{\kappa}{4}} e^{2\phi}[ (\rho')^{2} - \rho'' + t ]  .
\label{const}
\end{eqnarray}
where the prime denotes the derivative with respect to the coordinate 
$\sigma$. We shall impose the following boundary conditions 
at the horizon $\sigma \sim  -\infty$ which are necessary to have a black hole configuration with regular horizon. 
\begin{eqnarray}
    \rho(\sigma \sim -\infty)      &=& 
          \lambda \sigma + \rho_{0} 
       +  \rho_{1} e^{2\lambda\sigma}
       +  \rho_{2} e^{4\lambda\sigma} + \cdots  \label{bdy1}
                                                           \\ 
    \phi(\sigma \sim -\infty)      &=& 
          \phi_{0} 
       +  \phi_{1} e^{2\lambda\sigma}
       +  \phi_{2} e^{4\lambda\sigma} + \cdots \label{bdy2}
\end{eqnarray}
where,
\begin{eqnarray*}
    \rho_{0} &=& 0  \\
    \phi_{0} &=& \phi_{h} \\
    \rho_{1} &=& -\frac{\lambda^{2}}{2}
                  \frac{e^{2\phi_{h}}}{1-\frac{\kappa}{2} e^{2\phi_{h}}} \\
    \phi_{1} &=& \rho_{1} \\
    \rho_{2} &=& \frac{1}{4(1-\frac{\kappa}{2} e^{2\phi_{h}})} 
                 [ (1+\kappa e^{2\phi_{h}})(\phi_{1})^{2} 
                  - 2\lambda^{2} e^{2\phi_{h}}\phi_{1} ] \\
    \phi_{2} &=& \rho_{2} + \frac{1}{4}(\phi_{2})^{2}
\end{eqnarray*}
where $\rho_{0} = 0 $ is guaranteed by a scale transformation as 
in Lowe's paper. $\rho_{1},\rho_{2},\phi_{1} $ and 
$ \phi_{2} $ are obtained from equations of motion (\ref{sem1}-\ref{sem3}).
Using the above expansions one finds that the regularity condition for the 
solutions at the horizon will be 
\begin{eqnarray*}
    t &=& -\lambda^{2} .
\end{eqnarray*}
We are interested in the situation where there will be no naked singularity. 
Remembering that the dilaton field decreases monotonically\cite{L}, 
we will require the following condition to guarantee such situations.
\begin{eqnarray*}
    \phi_{cr}  >  \phi_{h}
\end{eqnarray*} 
We have performed numerical integration of the above set of equations  (\ref{sem1}-\ref{sem3}) with the boundary conditions (\ref{bdy1}) (\ref{bdy2}) 
to find static solutions for various parameters $(\phi_{cr}, 
\phi_{h})$.  The results coincide with the original calculations by Lowe.

\section{Perturbations around  static solutions}

In this section we shall study the linear perturbation of the static background solutions described in the previous section.
Thus we shall write our variables as 
\begin{eqnarray}
    \phi &=& \phi_{b}(\sigma) + \psi(x) \\
    \rho &=& \rho_{b}(\sigma) + h(x),   
\end{eqnarray}
where $ \phi_{b}(\sigma)$ and $ \rho_{b}(\sigma) $ are  static black hole solutions, and the perturbed variables $\psi$ and $h$ are assumed to be small 
in magnitudes. 

Then we obtain the equations for $\psi$ and $h$ from (\ref{sem1}) and (\ref{sem2}), respectively. 
\begin{eqnarray}
    \partial_{+}\partial_{-} \psi     &=& 
         - \frac{1}{2} {\phi'}_b
                       \frac{\partial \psi}{\partial\sigma} 
                                             \nonumber       \\
    \mbox{}                          &+&  
         \frac{1}{1 - \frac{\kappa}{2}e^{2\phi_b}}
         \left(  
         - \frac{1}{2} {\phi'}_b
                       \frac{\partial \psi}{\partial\sigma} 
     +   \lambda^{2} e^{2(\rho_b+\phi_b)}(\psi+h)
          \right)   
  \nonumber   \\
    \mbox{}                          &+&
         \frac{\kappa e^{2\phi_b}}{(1 - \frac{\kappa}{2}e^{2\phi_b})^{2}}
         \left(
          -\frac{{\phi_b}'^{2}}{4} + \frac{\lambda^{2}}{2} e^{2(\rho_b+\phi_b)}
         \right) \psi  
                                                                 \\
    \partial_{+}\partial_{-} h     &=& 
         \partial_{+}\partial_{-} \psi 
      +  \frac{1}{2} {\phi'}_b \frac{\partial \psi}{\partial \sigma} 
\end{eqnarray}     
The perturbed equation for the constraint equation (\ref{const}) may be 
also calculated as 
\begin{eqnarray*}
\delta <T_{\pm\pm}>    &=&
         e^{-2\phi} [ -   2 \partial^{2}_{\pm} \psi 
                     \pm  2 {\phi'}_b \partial_{\pm} h
                     \pm  2 {\rho'}_b \partial_{\pm} \psi
                     \pm  2 {\phi'}_b \partial_{\pm} \psi]
        -2 <T_{\pm\pm}>_b \psi
\end{eqnarray*}
where $<T_{\pm\pm}>$ is the trace anomaly part of the stress energy tensor. 
\begin{eqnarray}
<T_{\pm\pm}>      &=&
        \kappa[  (\partial_{\pm}\rho)^{2} 
       - \partial^2_{\pm}\rho
       + t_{\pm}(x^{\pm})], 
\end{eqnarray}
and $<T_{\pm\pm}>_b$ is the trace anomaly evaluated by the background solution $(\rho_b, \phi_b)$.

In the following analysis it turns out convenient to use 
the conformal gauge invariant quantities  for our perturbed variables.  
\begin{eqnarray}
\Phi &=& h - \frac{\partial}{\partial\sigma} 
                 \left[\frac{\psi}{{\phi'}_b}\right] 
               - \frac{d\rho}{d\phi}|_b \psi  \\
\Theta &=&  {\phi'}_b       
              \partial_{+}\partial_{-} \left[\frac{\psi}{{\phi'}_b} \right] \\
\Xi_{\pm}&=&  \delta <T_{\pm\pm}> 
            \mp 4\frac{<T_{\pm\pm}>_{b}}{{\phi'}_b}
            \partial_{\pm} \psi \nonumber   \\
    \mbox{}     &+&   
            \left(  2\frac{{\phi''}_b}{({\phi'}_b)^{2}} <T_{\pm\pm}>_{b} 
           - \frac{{<T_{\pm\pm}>'}_b}{{\phi'}_b} \right) \psi  
\end{eqnarray} 
Using the above variables, the perturbation equations 
take the following forms:
\begin{eqnarray}
    0  &=& 
           \partial_{+}\partial_{-}\Phi 
         +  \frac{\partial}{\partial\sigma}
            \left[\frac{\Theta}{{\phi'}_b}\right] 
         +  \frac{{\rho'}_b}{{\phi'}_b} \Theta - \Theta  \label{EQ1} \\
    0  &=& 
            \frac{\kappa}{2} e^{2\phi_b}
             \left(  \partial_{+}\partial_{-}\Phi 
                    + \frac{\partial}{\partial\sigma}
                      \left[\frac{\Theta}{{\phi'}_b}\right]
                    + \frac{{\rho'}_b}{{\phi'}_b} \Theta
             \right)
        +   \lambda^{2} e^{2(\phi_b+\rho_b)}\Phi -\Theta  \label{EQ2}\\
    \Xi_{\pm} &=& 
           -2 e^{-2\phi_b}(\Theta \mp {\phi'}_b\partial_{\pm}\Phi)  \label{EQ3}
\end{eqnarray}
The first two equations (\ref{EQ1})(\ref{EQ2}) give us 
\begin{eqnarray}
    \Theta   &=&  
             \frac{\lambda^{2} e^{2(\rho_b+\phi_b)}}
                  {1-\frac{\kappa}{2} e^{2\phi_b}} \Phi  
\end{eqnarray}
The last equation (\ref{EQ3}) may be formally solved as 
\begin{eqnarray}
    \Xi_{\pm} &=&  
                \pm 2{\phi'}_b e^{2\phi_b} 
                 \exp \left[2\int \frac{\Omega}{{\phi'}_b} d\sigma \right]
                 \partial_{\pm}\tilde{\Phi} 
\end{eqnarray}
where we have defined a new variables $\tilde{\Phi}$   
\begin{eqnarray}
    \Phi &=& 
         \exp \left[2 \int \frac{\Omega}{{\phi'}_b} d\sigma \right]
         \tilde{\Phi} 
\end{eqnarray}
with
\begin{eqnarray}
    \Omega &=& 
             \frac{\lambda^{2} e^{2(\rho_b+\phi_b)}}
                  {1-\frac{\kappa}{2} e^{2\phi_b}} 
\end{eqnarray}

Finally the first equation (\ref{EQ1}) becomes the wave equation with potential.
\begin{eqnarray}
    \left[  \frac{\partial^{2}}{\partial t^{2}} 
      - \frac{\partial^{2}}{\partial \sigma^{2}}
      + V[\rho_{b}(\sigma),\phi_{b}(\sigma)] \right] \tilde{\Phi}(x)
     &=& 0 . 
\end{eqnarray} 
where the potential is written as follows.
\begin{eqnarray}
    V &=& \frac{2\kappa\lambda^{2}e^{2\rho_{b}+4\phi_{b}}}
               {(\phi'_{b})^{2}(1-\frac{\kappa e^{2\phi_{b}}}{2})^{2}}
          \left[  (\phi'_{b} - \rho'_{b})^{2} + \frac{(\phi'_{b})^{2}}{2}
                - \lambda^{2}
          \right]
\end{eqnarray}
Unstable mode is obtained by solving the stationary solutions for this 
equation.
\begin{eqnarray*}
    \left[ - \frac{\partial^{2}}{\partial \sigma^{2}}
           + V[\rho_{b}(\sigma),\phi_{b}(\sigma)] \right] \tilde{\Phi}(x)
     &=& E \tilde{\Phi}(x) . 
\end{eqnarray*}
where $ E < 0 $ for unstable mode.

The asymptotic form of the potential near the horizon is evaluated as  
\begin{eqnarray}
    V         &=& 
          \frac{4\kappa e^{4\phi_{h}} \phi_{2}}
               {3 \phi_{1} (1-\frac{\kappa}{2} e^{2\phi_{h}})^{2}}
          e^{4\lambda\sigma}  \nonumber  \\
    \mbox{}   &=& 
         - \frac{\kappa\lambda^{2} (1-\frac{\kappa}{4} e^{2\phi_{h}})
                 e^{6\phi_{h}}} 
                {(1-\frac{\kappa}{2} e^{2\phi_{h}})^{4}}
            e^{4\lambda\sigma} < 0   
\end{eqnarray}
Thus the potential becomes negative near the horizon. 
It will be shown by numerical calculation in the next section that 
the potential is negative over an entire range of the region. 
This fact immediately leads  to the existence of at least one 
unstable mode.

\section{Numerical results}

In this section we shall explicitly solve our basic equations numerically 
and show that there is at least one unstable mode in Lowe's model 
described by the action (\ref{action}).

Unlike Lowe'paper, all the calculations have been done in the tortoise coordinate $\sigma$, since we find it is much convenient to see 
the behaviour of the potential and its dependence on the background
solutions.
 
Our results are summarized as follows.

\begin{description}
\item[1.] The potential is negative definite over the entire 
region of $\sigma$ for all region of parameters $(\phi_{cr}, \phi_h)$
that we have calculated. ($-2.5 < \phi_{cr} < 0.0,-2.501 < \phi_{h} < 
-0.02 $)
\item[2.] Although we can not prove analytically, our numerical results 
suggest that the shape of the potential depends only on the combination 
of $\phi_{cr} - \phi_{h} $.
\end{description}
Result 1 and well-known theorem\cite{LL} 
show that there will be at least one unstable  mode at least in the parameter region we have calculated.
Fig. 1 shows the potential as well as bound state. Fig. 2 shows 
also some potentials for a different choice of parameters, 
but the same combination $\phi_{cr} - \phi_{h} $.

\section{Summary}

We have investigated the dynamical instability of a quantum black hole in  two-dimensional dilaton gravity. The unperturbed solutions are taken as 
the static solutions describing a black hole with quantum correction 
and with incoming flux of radiation. The quantum backreaction is essential to have such static configurations. We have found a surprising result 
that static solutions found by Lowe in his model are dynamically unstable for linear perturbations.  This will mean that Lowe's model is not appropriate 
for the study of black hole evaporation unless some nonlinear mechanism 
to stabilize the solution is present.It can be also shown that 
the static solutions generated also by quantum backreaction in CGHS model 
are dynamically unstable. 
These examples strongly suggest that the stability analysis of quantum black 
hole model should be performed for each model of two-dimensional dilaton 
gravity, before withdrawing any physical conclusions using such models. 
In this sense it is interesting to see the situation in more general two-dimensional models \cite{FHI} which allows to have black hole configurations.  Particularly it would be interesting to 
investigate the stability of the models derived entirely from higher-dimensional theories by Busso and Hawking\cite{BH}. 
Finally it should be mentioned that our study opens up also 
a new interesting possibility to have some 
stable bounded modes outside the horizon for certain models of two-dimensional 
dilaton gravity. 
This work is now under investigation.

%\section*{Acknowledgment} 

\
\\

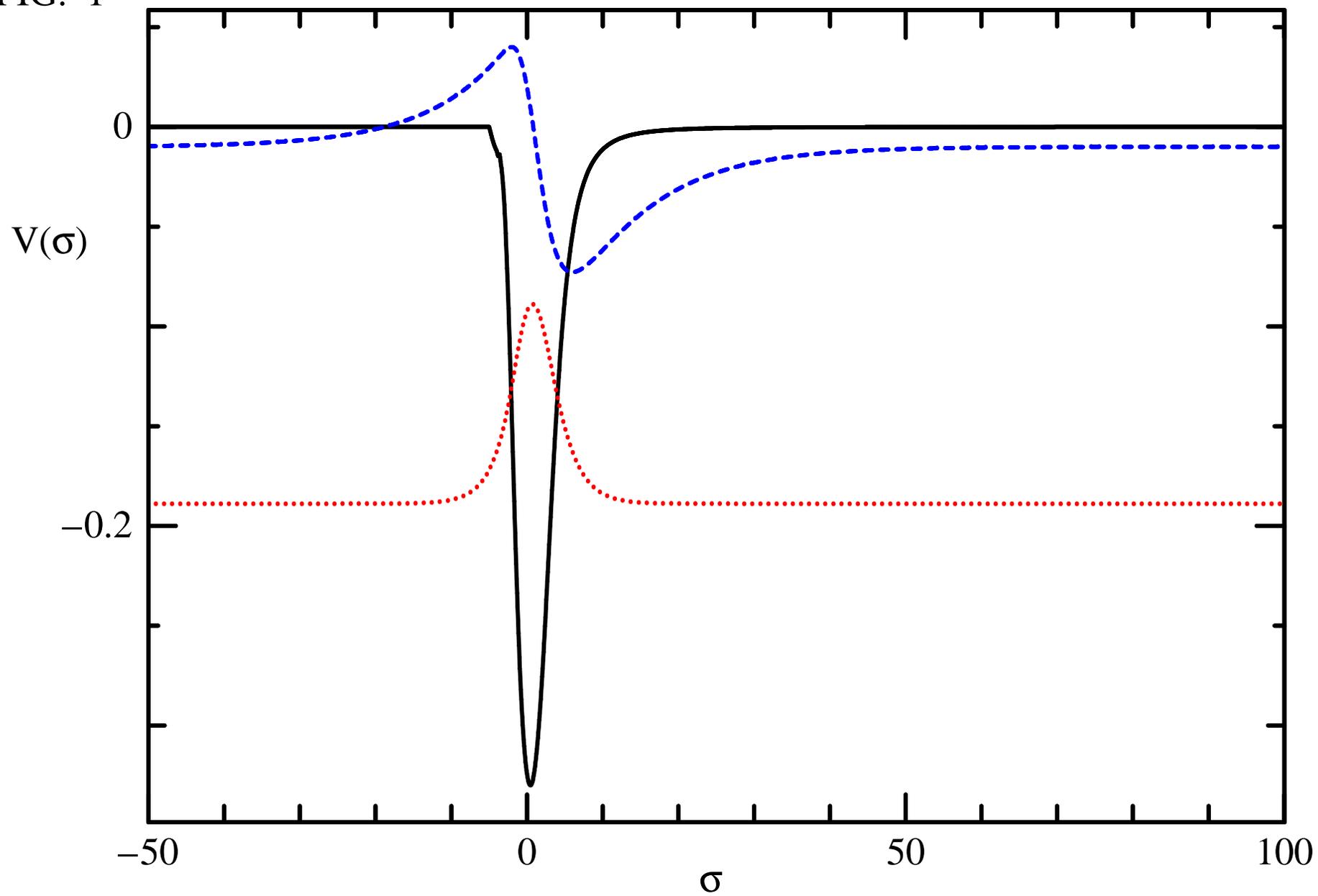
FIG. 1.
Potential and bound states are plotted in the radial coordinate $\sigma$ 
,where  parameters are  chosen as $\phi_{cr} = 1.5$, $\phi_{h}=1.52$ and 
$\lambda=1/\sqrt{2}$.
Here,solid line for potential ,dotted line for groundstate and broken
line for first excited state.\\

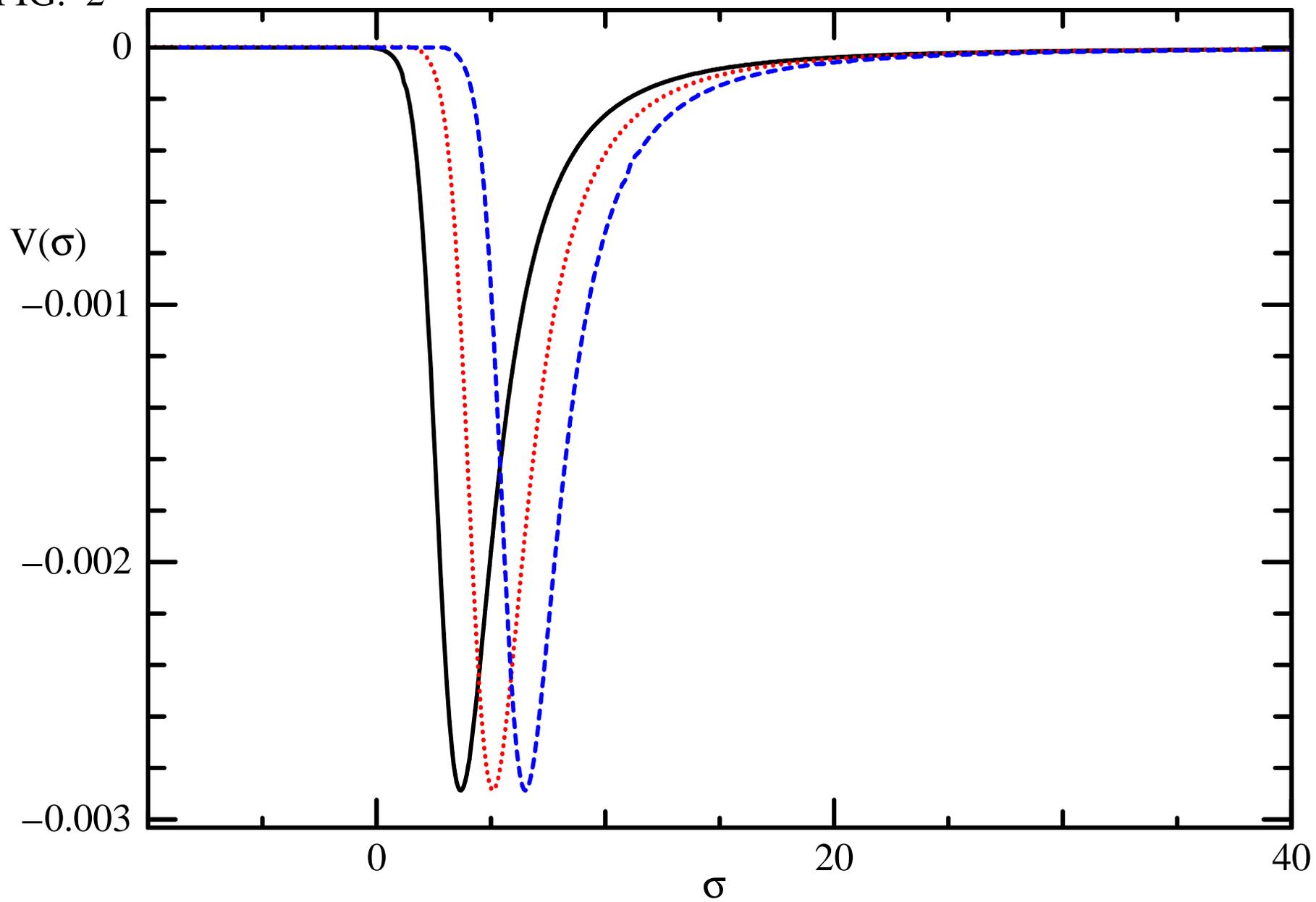
FIG. 2. 
Some potentials are plotted in the radial coordinate
$\sigma$ with $\lambda=1/\sqrt{2}$. 
Parameters ($\phi_{cr},\phi_{h}$) are chosen as $(0.0,2.02)$(solid
line),
$(1.0,3.02)$(dotted line) 
and ($2.0,4.02)$(broken line).

\end{document}